\documentclass[%
reprint,
 amsmath,amssymb,
 aps,
]{revtex4-1}

\usepackage{graphicx}
\usepackage{dcolumn}
\usepackage{bm}

\begin{document}

\title{Shapes of the $^{192,190}$Pb ground states from beta decay studies using the total absorption technique}

\author{M. E. Est\'evez Aguado}
\author{A. Algora}
\altaffiliation[On leave from ]{Institute of Nuclear Research, Debrecen, Hungary}%
\email{algora@ific.uv.es}
\author{J. Agramunt}
\affiliation{IFIC (CSIC - Universidad de Valencia), Valencia, Spain}
\author{B. Rubio}
\affiliation{IFIC (CSIC - Universidad de Valencia), Valencia, Spain}
\author{J. L. Ta\'in}
\affiliation{IFIC (CSIC - Universidad de Valencia), Valencia, Spain}
\author{D. Jord\'an}
\affiliation{IFIC (CSIC - Universidad de Valencia), Valencia, Spain}
\author{L. M. Fraile}
\affiliation{Universidad Complutense de Madrid, CEI Moncloa, 28040 Madrid, Spain}
\author{W. Gelletly}
\affiliation{University of Surrey, Guildford, UK}
\author{A. Frank}
\affiliation{Instituto de Ciencias Nucleares, UNAM, M\'exico}
\author{M. Csatl\'os}
\address{Institute of Nuclear Research, Debrecen, Hungary}
\author{L. Csige}
\address{Institute of Nuclear Research, Debrecen, Hungary}
\author{Zs. Dombr\'adi}
\address{Institute of Nuclear Research, Debrecen, Hungary}
\author{A. Krasznahorkay}
\address{Institute of Nuclear Research, Debrecen, Hungary}
\author{E. N\'acher}
\address{Instituto de Estructura de la Materia, CSIC, Madrid, Spain}
\author{P. Sarriguren}
\address{Instituto de Estructura de la Materia, CSIC, Madrid, Spain}
\author{M. J. G. Borge}
\address{Instituto de Estructura de la Materia, CSIC, Madrid, Spain}
\author{J. A. Briz}
\address{Instituto de Estructura de la Materia, CSIC, Madrid, Spain}
\author{O. Tengblad}
\address{Instituto de Estructura de la Materia, CSIC, Madrid, Spain}
\author{F. Molina}
\address{Comision Chilena de Energia Nuclear, Santiago de Chile, Chile}
\author{O. Moreno}
\address{Massachusetts Institute of Technology, Cambridge, MA 02139, USA}
\author{M. Kowalska}
\address{CERN, Switzerland}
\author{V. N. Fedosseev}
\address{CERN, Switzerland}
\author{B. A. Marsh}
\address{CERN, Switzerland}
\author{D. V. Fedorov}
\address{Petersburg Nuclear Physics Institute, Gatchina, Russia}
\author{P. L. Molkanov}
\address{Petersburg Nuclear Physics Institute, Gatchina, Russia}
\author{A. N. Andreyev}
\address{University of York, York, UK}
\author{M. D. Seliverstov}
\address{IP, University of Mainz, Germany}
\author{K. Burkard}
\address{GSI, Darmstadt, Germany}
\author{W. H\"uller}
\address{GSI, Darmstadt, Germany}

\begin{abstract}

The beta decay of $^{192,190}$Pb has been studied using the total absorption technique at the ISOLDE(CERN) facility. The beta-decay strength deduced from the measurements, combined with QRPA theoretical calculations, allow us to infer that the ground states of the $^{192,190}$Pb isotopes are spherical. These results represent the first application of the shape determination method  using the total absorption technique for heavy nuclei and in a region where there is considerable interest in nuclear shapes and shape effects. 
\end{abstract}

\pacs{23.40.Hc,27.80.+w,29.30.Kv}
\keywords{total absorption \sep beta decay \sep ground state nuclear shape determination \sep proton rich lead isotopes } 

\maketitle

\section{Introduction}
\label{intro}


 One peculiarity of nuclear systems is that in general they are not spherical and take up quite a variety of different shapes. They may have axially deformed prolate and oblate, triaxial, and octupole shapes  \cite{nature_butler, Nyako,Chandler,nazarewicz}. As a result the description of nuclear shapes presents a major challenge to nuclear models. Here we are concerned with shapes in heavy nuclei and in particular with the determination of shapes using total absorption spectroscopy. 
The method, which is based on a proper measurement of the beta strength as a function of excitation energy in the daughter nucleus and its comparison with theoretical calculations, has only been applied up to now to medium-mass nuclei in the $A \sim 80$ region \cite{Nacher2004,Poirier2004,Perez2013}. 
The present result illustrates the more general scope of the total absorption technique as a tool for inferring the shapes of exotic nuclei when combined with the results of shape-sensitive theoretical calculations.
In this study we have selected the neutron-deficient Pb nuclei. They have been the subject of intense experimental and theoretical interest in recent years. The main reason is the existence of at least one $0^+$ excited state below 1 MeV in each of the even-even Pb isotopes between A=184 and A=194. This offers excellent opportunities to study shape effects and shape co-existence \cite{Julin2001}. The cases that have attracted most interest are the $^{186,188}$Pb nuclei where there are even two $0^+$ excited states below 700 keV \cite{Andreyev2000,Heese1993}. Theoretically, the existence of the  low-lying excited $0^+$  states has been interpreted as the result of the combined effect of the proton shell gap at Z=82 and the influence of a large number of neutron holes below N=126. The role of the magic proton number Z=82 is considered to be determinant in the structure of these nuclei.  All of them, down to $^{182}$Pb, are expected to be spherical in their ground states \cite{Witte2007}.

Theoretically, the co-existence of the low-lying excited $0^+$ states has been studied in the framework of a variety of models \cite{Heyde2011}. In a shell model picture, the excited $0^+$ states are interpreted as being due to two-quasiparticle and four-quasiparticle configurations \cite{Heyde1991}. On the other hand calculations based on phenomenological mean field models and the Strutinsky method predict the existence of several competing minima in the deformation surfaces of these nuclei \cite{Bengtsson1989,Nazarewicz1993}. Self-consistent mean field calculations \cite{Smirnova2003,Niksic2002} and calculations including correlations beyond the mean field \cite{Duguet2003,Egido2004,Rodriguez2004,Bender2004} confirm these results. This problem has also been studied in the framework of the interacting boson model \cite{Fossion2003,Frank2004}. 

The basis of the present work lies in the fact that it has been shown theoretically that the beta decay properties of unstable nuclei may depend on the shape of the decaying parent nucleus \cite{Hamamoto1995,Frisk1995,Sarriguren1998,Sarriguren1999,Sarriguren2001}. In particular cases the calculated Gamow-Teller strength distribution shows different patterns depending on the shape of the parent nucleus \cite{Sarriguren1998,Sarriguren1999,Sarriguren2001}. This property can be used to infer the shape of the ground state of the decaying nucleus if the Gamow-Teller strength distribution has been measured properly. For this type of experiment the total absorption spectroscopy (TAS) technique is needed, which is the only technique that can provide beta intensity distributions free from the pandemonium effect \cite{hardy1977}. Measurements performed at ISOLDE of the beta decays of the neutron deficient $^{74}$Kr and $^{76,78}$Sr nuclei using the total absorption spectrometer {\it Lucrecia} have shown the potential of this method \cite{Nacher2004,Poirier2004,Perez2013} for nuclei in the $A \sim 80$ region. 

Theoretical calculations \cite{Sarriguren2005,Moreno2006} predict that the Gamow-Teller strength distributions in the decay of the even-even $^{184,186,188,190,192,194}$Pb isotopes have quite differing patterns depending on the  deformation assumed for the parent state. This feature can be used to obtain new insight into the shape co-existence phenomenon in this region if the theoretical results are combined with accurate measurements of the B(GT) distributions in these nuclei. This was the primary motivation for the present work. In addition,  precise measurements of the B(GT) distributions in these nuclei are also important to test nuclear models further in the Z=82 region. The decays of  $^{190,192}$Pb are of particular interest in this context, since the degree of shape mixing in the parent ground state is expected to be small \cite{Fossion2003}.


\section{The Experiment}
\label{experiment}

The measurements were performed at ISOLDE (CERN). A  50 gcm$^{-2}$ UC$_x$ target was bombarded with a beam of 1.4 GeV protons from the Proton Synchrotron Booster (PSB) to produce
the radioactive isotopes of interest. The selective laser ionisation source, RILIS \cite{RILIS},  was used to enhance the ionisation of Pb isotopes compared to their Tl isobars,
which are surface ionised. Then the General Purpose Separator (GPS) was employed to achieve separation in mass ($A$). Two measurements were made for each mass: one with laser
on, to produce a radioactive beam of the lead isotope of interest, and one with laser off, to produce a beam dominated by Tl to study possible contaminantion in the Pb measurements.
The purities of the $^{192}$Pb and $^{190}$Pb beams with the laser-on were determined to be 96 \% and 93 \% respectively. In this work we present the results of the analysis of the laser-on data for the beta decays of the $^{192}$Pb and $^{190}$Pb isotopes. 

Once produced and separated, the beam of the ions with the selected mass number was transported to the total absorption spectrometer measuring station {\it Lucrecia}. The beam was implanted in a magnetic tape outside the setup, that was then moved in a single step to the centre of the TAS detector for the measurements. One measuring cycle was composed of symmetrical collection and measuring times, which were adjusted to reduce the impact of daugther decay contamination in the measurements (cycle lengths: 4 min. for $^{192}$Pb and 2 min. for $^{190}$Pb), after which the tape was moved again. The process was repeated until sufficient data were recorded.

The TAS detector at ISOLDE, \textit{Lucrecia}, is made of a  cylindrically shaped NaI(Tl) mono-crystal with $\phi = h = $ 38 cm (see the schematic figure of the setup in Ref.
\cite{Perez2013}). The detector has a transverse hole of $\phi =7.5$ cm perpendicular to its symmetry axis. The beam pipe is inserted in this hole and it is positioned so that the source is carried to the centre of the detector for the measurements.  On the side opposite to the beampipe, ancillary detectors can be placed for the measurement of X rays, $\gamma$ rays, betas, etc. in coincidence with the TAS detector.  The scintillation light from \emph{Lucrecia} is collected by 8 photomultipliers. The total efficiency of the detector is around 90 \% for mono-energetic gamma-rays from 300 to 3000 keV. During the measurements the counting rate in \emph{Lucrecia}  was kept below 10 kHz to reduce the effect of second and higher order pileup contributions.  \emph{Lucrecia} had an energy resolution of 7.5 $\%$ at the 662 keV $^{137}$Cs peak.  

In the measurements a germanium telescope was used as an ancillary detector. The telescope is composed of a Ge planar and a Ge coaxial detector, with the planar detector positioned closer to the tape.  The planar detector was used with two energy gains: one optimised for the measurement of the X-rays emitted in the electron capture (EC) decay and the other for detecting low-energy gamma rays. The coaxial detector was used for the measurement of the individual gammas emitted in the de-excitation cascades and for looking at gamma-gamma coincidences with the planar detector. 

\begin{figure}[b]
   \begin{center}
  \includegraphics[width=6.5cm]{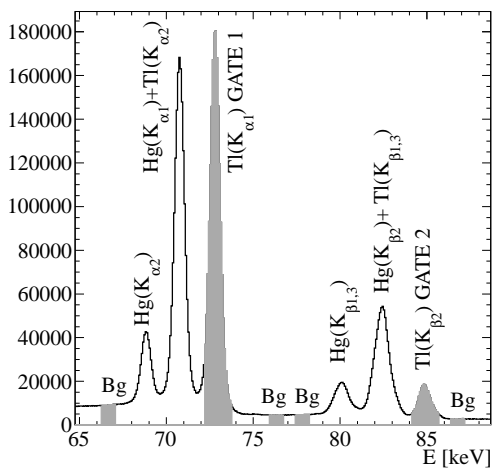}
    \caption{X-ray gates used for the generation of the analysed spectrum for $^{192}Pb$ decay. The gates are highlighted in grey color. Gates marked with "Bg"
    represent the background gates used for subtraction. }\label{fig:Xrays192}
    \end{center}
  \end{figure}

\begin{figure}[b]
   \begin{center}
    \includegraphics[width=6.5cm]{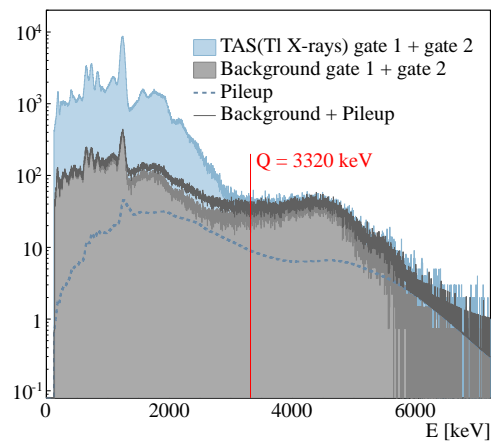}
    \caption{Contributions of the different spectra for the analysed spectrum of $^{192}Pb$ decay. See Fig. 1 for details on the gates.}\label{fig:spectra192}
    \end{center}
  \end{figure}

\section{Results}

For the analysis of the two decays presented below, a TAS gated spectrum was generated by requesting coincidences between the TAS and the Tl X-rays in the planar detector (EC component of the Pb decay).  
Comparison of the gates set for different X-ray lines of Tl and Hg showed different patterns. This shows the cleanliness of the generated X-ray gated TAS spectra (see
Figs.\ref{fig:Xrays192} and \ref{fig:spectra192}). In addition, the $\beta^+$ component of the beta decay in the X-ray gated spectra was estimated to be negligible. Accordingly its
presence in the X-ray gates was discounted. 

The X-ray gated TAS spectra were generated by putting gates on the X-ray peaks and subtracting the background using gates set on the continuous background spectrum as shown in  Figs.\ref{fig:Xrays192}.  Thus the background and pileup, which distort the spectra were taken into account in the generation of the spectra for analysis (see Fig.\ref{fig:spectra192} for the 192 case).


To analyse the data we need to solve the TAS inverse problem:

\begin{equation}\label{eq:drf}
  d_i = \displaystyle\sum_{j=0}^{j_{max}}R_{ij}(B)f_j
\end{equation}

where $d_i$ is the content of bin $i$ in the contaminant-free TAS spectrum, $R_{ij}$ is the response matrix of the TAS setup and represents the probability that a decay that feeds level $j$ in the daughter nucleus gives a count in bin $i$ of the TAS spectrum, and $f_j$ is the beta feeding to the level $j$. The response matrix $R_{ij}$ depends on the TAS setup and on the assumed level scheme of the daughter nucleus ($B$). To calculate the response matrix first the branching ratio matrix $B$ for the levels in the daughter nucleus has to be determined. For that the level scheme of the daughter nucleus is divided into two regions, a low excitation part and a high excitation part. Conventionally the levels of the low excitation part and their branchings are taken from high resolution measurements available in the literature, since it is assumed that the gamma branching ratios of these levels are well known. Above a certain energy, the cut energy, a continuum of possible levels divided in 40 keV bins is assumed.  From this energy up to the  $Q_{EC}$ value, the statistical model is used to generate a branching ratio matrix for the high excitation part of the level scheme. The statistical model is based on a Back-Shifted Fermi Gas Model level density function \cite{dilg1973} and gamma strength functions of E1, M1, and E2 character. The parameters for the gamma strength function were taken from \cite{capote2009} 
and the parameters of the level density function were obtained from fits to the data available in 
\cite{goriely2001,demetriou2001,capote2009}. As part of the optimisation procedure in the analysis, the cut energy and the parameters of the statistical model can be changed. Once the branching ratio matrix is defined, the $R_{ij}$ can be calculated recursively from previously calculated responses using Monte Carlo simulations \cite{cano1999monte,cano1999pulse,agostinelli2003}. The Monte Carlo simulations were validated with measurements of the spectra of well known radioactive sources ($^{24}$Na, $^{60}$Co, $^{137}$Cs).  Once the $R$ response matrix is obtained, the Expectation Maximisation (EM) algorithm is applied to extract the beta feeding distributions \cite{tain2007influence,tain2007algorithms}. 

\begin{figure}[b]
    \begin{center}
    \includegraphics[width=6.5cm]{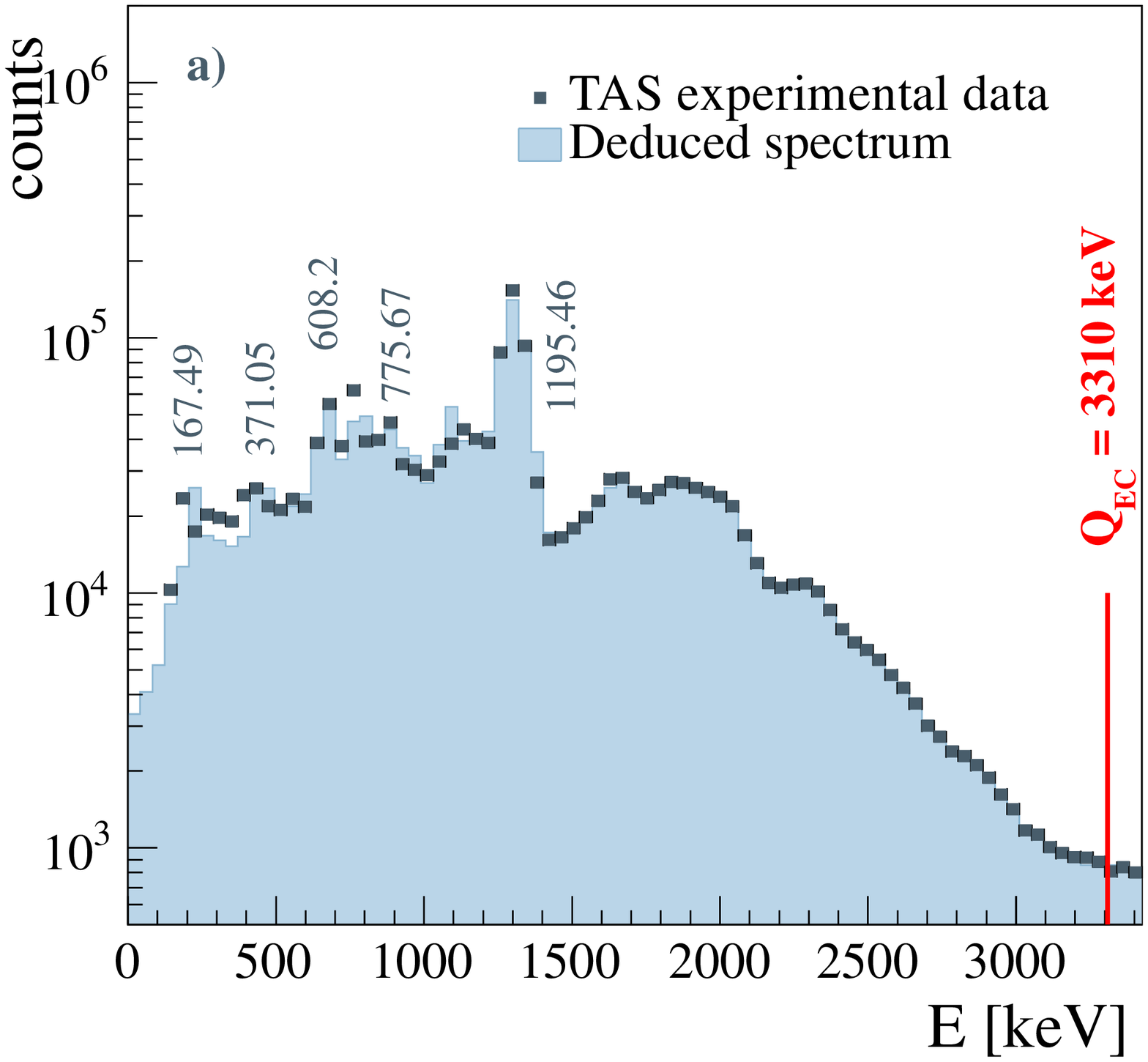}\hspace{5mm}\includegraphics[width=6.5cm]{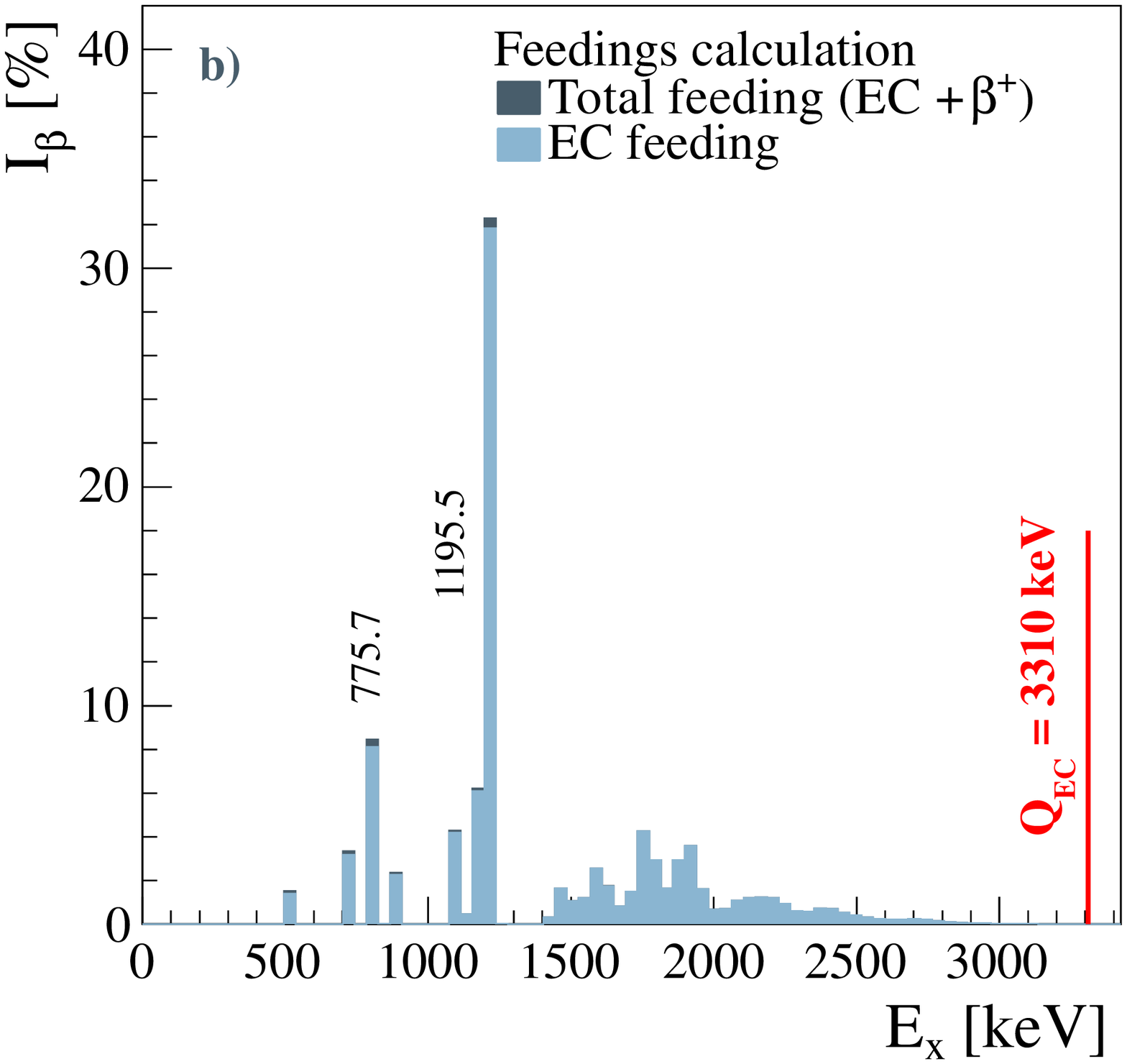}
    \caption{Comparison of the analyzed and reconstructed spectrum after the analysis for the $^{192}Pb$ decay (a) and the deduced feeding distribution
    (b).}\label{fig:bayes192}
    \end{center}
  \end{figure}

\begin{figure}[b]
    \begin{center}
    \includegraphics[width=6.5cm]{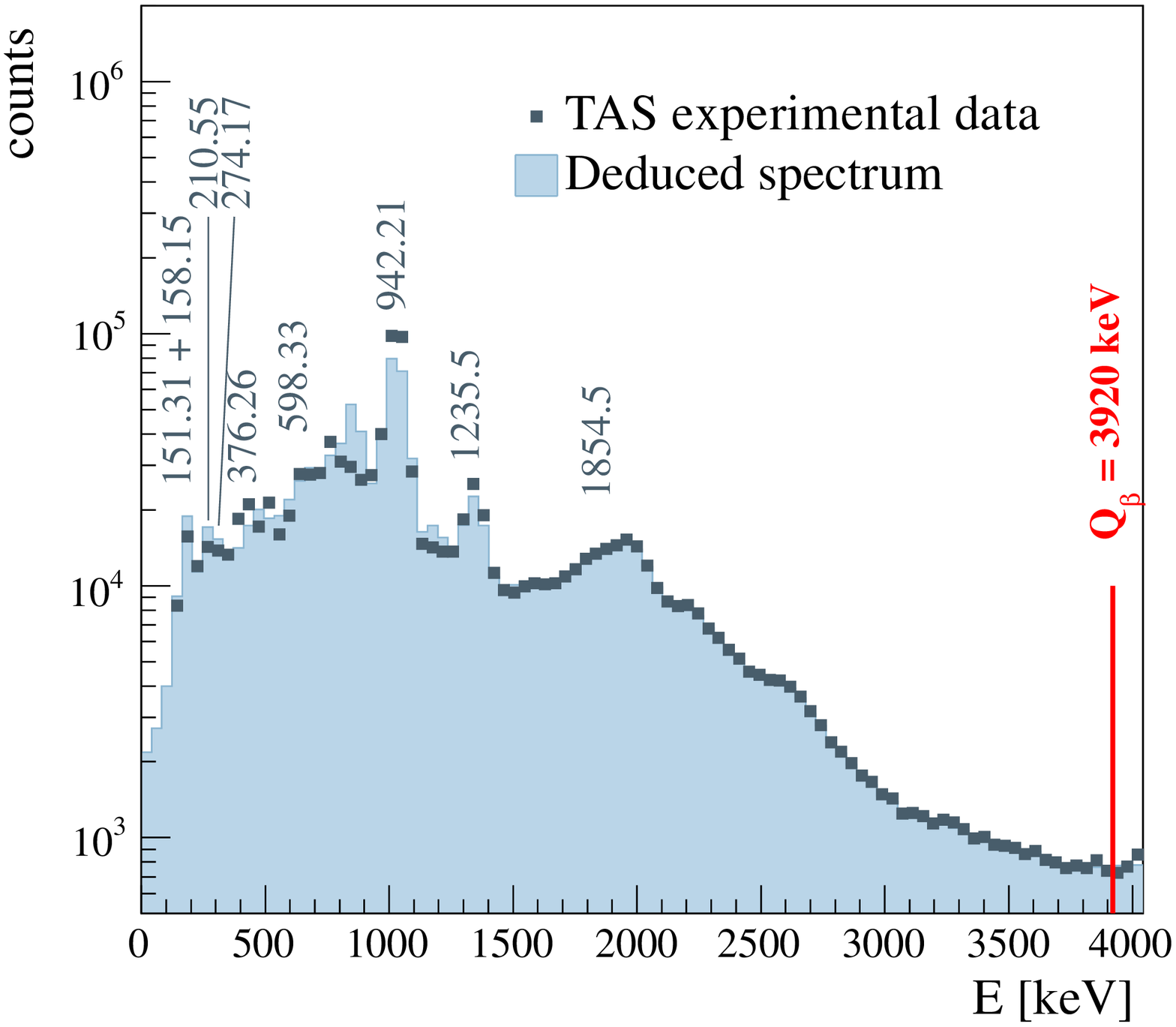}\hspace{5mm}\includegraphics[width=6.5cm]{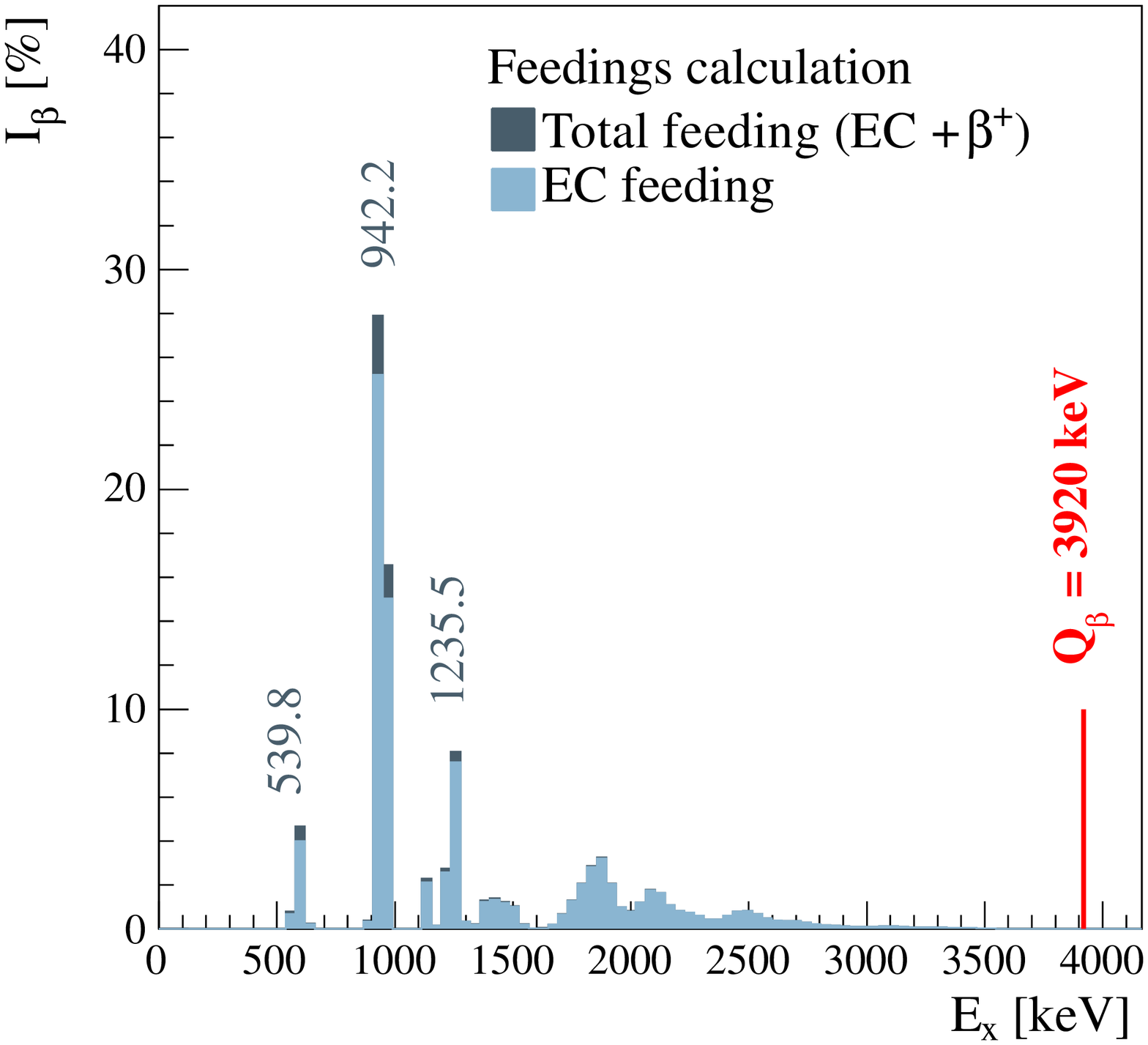}
    \caption{(a) Comparison of the analyzed and reconstructed spectrum after the analysis for the $^{190}Pb$ decay and (b) the feeding distribution deduced from the analysis.
    }\label{fig:bayes190}
    \end{center}
  \end{figure}

The decay level scheme taken from high-resolution data was found to be incomplete for both of the cases presented in this work. The level scheme of $^{192}$Tl had to be
modified in order to reproduce the measured TAS spectrum. Two levels were added at 494 keV and 690 keV, based on two gammas seen in the germanium detector and on two
unexplained peaks appearing in the X-ray gated TAS spectrum at those energies. When these levels were added to the level scheme, the measured spectrum was correctly
reproduced in the analysis. In this case the cut energy for the statistical model was placed at 800 keV in the calculation of the branching ratio matrix. A comparison of the
analysed TAS spectrum with that reconstructed after the analysis is presented in Fig. \ref{fig:bayes192}. For the analysis of the decay of $^{190}$Pb the cut energy for the
application of the statistical model in the level scheme of $^{190}$Tl was placed at 520 keV after several trials for this parameter. With this value, a similar quality to
the 192 case was obtained in the reproduction of the analysed spectrum as can be seen in Fig.\ref{fig:bayes190}.\\

The quantity of interest for this study is the beta strength. The experimental strength $S_{\beta}$ was calculated from the beta intensities per energy bin $I_{\beta}$ according to Eq. \ref{eq:sexp}:

\begin{equation}\label{eq:sexp}
  S_\beta(E) = \frac{I_\beta(E)}{f(Q_{EC} - E)t_{1/2}}
\end{equation}

where $E$ is the excitation energy in the daughter nucleus, $Q_{EC}$ and $t_{1/2}$ are the $Q$   value and half-life of the decay and $f(Q_{EC} - E)$ is the statistical rate function. 
Our analysis based on the X-ray gated TAS spectra provides the EC component of the decay. The $I_{\beta}$ ($I_{EC}$+ $I_{\beta^{+}}$) was calculated using the $I_{EC}$/$I_{\beta^{+}}$ tables of Ref. \cite{Gove}.

In the strength calculation for the beta decay of $^{192}$Pb into $^{192}$Tl the $Q_{EC}$ value
of 3310(30) keV \cite{wang2012} and $t_{1/2}$ = 3.5(1) min \cite{Sousa1981} were used. The EC+
$\beta^{+}$ branch for this decay is 99.9941(7) \% \citep{baglin2012,Sousa1981} (the rest is due to alpha decay). For the beta decay of $^{190}$Pb into $^{190}$Tl, the values of
$Q_{EC}$ = 3960(50) \cite{wang2012} and $t_{1/2}$ = 71(1) s \cite{Richards1996} were employed. In
this case the EC+ $\beta^{+}$ branch is 99.60(4) \cite{Singh2003,Ellis1981} and as in the
$^{192}$Pb case, the remainder of the decay goes is by alpha emission. In Tables \ref{tab:table1}
and \ref{tab:table2} the results for the beta feedings from the measurements and the deduced beta strength are shown.

\begin{table*}
\caption{\label{tab:table1}  Deduced $I_{EC}$ and $I_{EC+\beta^{+}}$ values ($I_{\beta}$ values) for the levels populated in the decay of $^{192}Pb$ for the accepted analysis. The last row of the table (total) includes the sum for all the levels up to the last accepted level in the analysis. Sums for two energy intervals are also given for the TAS results, one from the cut energy up to the last accepted level in the analysis and the other from the last known level from high resolution up to the last accepted level. 
The high-resolution (HR) data were taken from \cite{baglin2012,Sousa1981} and the $Q_{EC}$ value from \cite{wang2012}. 
}
\begin{ruledtabular}
\begin{tabular}{r c r r r r r }
        $E_{lev}$ & $J^{\pi}$	& \multicolumn{2}{c}{$I_{EC}$ [\%]}	& \multicolumn{2}{c}{$I_{EC+\beta^{+}}$ [\%]}	& \multicolumn{1}{c}{$B$(GT) [$g_A^2/4\pi$]} \\
	 $[keV]$	 &		& $TAS$		     & $HR$		& $TAS$			& $HR$		        & $TAS$			\\ \hline
	167.49   & 1($^{-}$)	         & $\sim$0 	     & 13(9)		& $\sim$0		& 15(10)		& 		         \\
	371.05   & 1($^{-}$)	         &  $\sim$0	     & 5.8(14)	& $\sim$0		& 6.3(15)		& $\sim$0  	\\
	413.98   & (1$^{-}$,2$^{-}$)	&  $\sim$0     & -		& $\sim$0  	& -			& $\sim$0	         \\
        494 \footnote{Levels added to the known level scheme to reproduce the experimental TAS spectrum.}  &                              &  1.4(8)	     & -		& 1.5(9)	         & -			& 0.001(1)	         \\
        690 \footnotemark[1]  &   	                  &  3.2(22)	     & -		& 3.3(2.4)		& -			& 0.003(2)	         \\
	775.67   & (0$^{-}$,1$^{-}$) &  8.1(32)	     & 22(3)		& 8.4(33)		& 23(3)		& 0.007(5)	         \\
	1195.46 &	1$^{+}$ 	         & 38(4)	     & 57(7)		& 38(4)		& 58(7)		& 0.05(2)	         \\	
    800-3200  &	   	                  & 86(5)	     & -	         & 87(6)		& -		         & 0.24(4)	         \\
  1240-3200  &	   	                  & 41(2)	     & -	         & 41(2)		& -		         & 0.18(3)	         \\ \hline	
	TOTAL     &		                 & 98(7)	     & 98(9)		& 100(7)		& 102(1)		& 0.24(4)		
\end{tabular}
\end{ruledtabular}
\end{table*}


\begin{table*}
\caption{\label{tab:table2} Same as Table \ref{tab:table1} but for the  $^{190}Pb$ decay case. The HR data were taken from \cite{Singh2003,Ellis1981} and the $Q_{EC}$ value from \cite{wang2012}. },
\begin{ruledtabular}
\begin{tabular}{r c r r r r r }

      $E_{lev}$ & $J^{\pi}$	 & \multicolumn{2}{c}{$I_{EC}$ [\%]}	 & \multicolumn{2}{c}{$I_{EC+\beta^{+}}$ [\%]}   & \multicolumn{1}{c}{$B$(GT) [$g_A^2/4\pi$]}\\
	  $[keV]$ &		 & $TAS$  	 & $HR$			 & $TAS$  		 & $HR$			 & $TAS$  			 \\ \hline
       151.31 \footnote{Levels not resolved by the TAS, their $I_{\beta}$ is assigned to levels 151, 376, 540 and 942 respectively.}	& (1)$^-$		   & $\sim$ 0	 & $<$ 17		 & $\sim$ 0		 & $<$ 21		 & $\sim$ 0	 \\
       158.15 \footnotemark[1]	& (0,1)$^-$		   &			  & 2.1(12)		 &			 & 2.6(15)		 &		       \\
       210.55	     & (1)$^-$  	   & $\sim$ 0	 & $<$ 17		 & $\sim$ 0		 & $<$ 21		 & $\sim$ 0		\\
       274.17	     & (1,2)$^-$	   & $\sim$ 0	 &  * \footnote{No information available}			 & $\sim$ 0		 & *
       \footnotemark[2] 	 & $\sim$ 0		\\
       372.75\footnotemark[1]	& (0$^-$, 1$^-$)  &			   & 1.3(5)		 &			  & 1.6(6)		 &			 \\
       376.26\footnotemark[1]	& (1)$^-$		  & $\sim$ 0		  & 4.6(16)		& $\sim$ 0	   & 5.4(19)		 & $\sim$ 0	   \\
       416.68	     & -		  & 0.6(3)		 & $<$ 19		 & 0.7(4)		  & $<$ 22		 & 0.0008(4)	    \\
       495.07\footnotemark[1]	& (1$^-$)		  &			    &$<$ 19		 &			   & $<$ 22		 &			  \\
       539.81\footnotemark[1]	& (0,1) 		  & 1.1(5)		   & 1.7(4)		 & 1.3(5)		 & 2.0(4)		 & 0.002(1)		  \\
       598.33	     & (1$^-$)  	  & $\sim$ 0		 & $<$ 9.7		 & $\sim$ 0		 & $<$ 11		 & $\sim$ 0		 \\
       738.99	     & (0$^-$, 1)		  & $\sim$ 0		  & $<$ 9.9		 & $\sim$ 0		 & $<$ 11		 & $\sim$ 0		 \\
       890.72\footnotemark[1]  & (0$^-$, 1)		  &			   & 3.8(6)		 &			 & 4.1(6)		 &			   \\
       942.21\footnotemark[1]  & 1$^+$ 	  & 43.9(10)		  & 38(5)		 & 48.5(11)		 & 41(5)	 & 0.081(4)		 \\
       1235.5	     & 1		  & 10.6(3)		  & 4.3(6)		 & 11.3(4)		 & 4.5(6)		 & 0.024(2)		 \\
       1854.5	     & 1		  & 5.9(9)		  & 0.7(2)		 & 6.0(12)		 & 0.7(2)		 & 0.022(2)		 \\
     520-3700	 & -			  & 92(1)		  & -			 & 98(1)		 &  -			 & 0.32(4)		 \\
    1880-3700	& -			  & 18(1)		& -			 & 18(1)	 &  -			 & 0.14(3)	                  \\ \hline 
 	 TOTAL  &				  & 94(2)		& 99.6(4)		   & 100(2)		  &  100.0(4)	 & 0.32(4)

\end{tabular}
\end{ruledtabular}
\end{table*}

%
%


\section{Discussion and Conclusions}

The primary aim of this work was to compare the beta strength distribution obtained from the TAS measurements with the calculations of Refs. \cite{Sarriguren2005, Moreno2006}, in order to infer the shape of the parent nucleus. The TAS results indicate that the high resolution studies of these decays suffered from the pandemonium effect \cite{hardy1977}.  The measurements showed a redistribution of the beta decay probability compared to earlier high resolution measurements, and a sizable amount of feeding up to the $Q_{EC}$ value, which is characteristic of this systematic error. 
More specifically, the last level observed in high resolution studies of the decay of $^{192}Pb$ was at 1195 keV and above that energy we have detected 41 \% feeding. Similarly in $^{192}Pb$ the last level observed in high resolution was at 1854 keV and we have detected 18 \% feeding from that level up to the $Q_{EC}$ value. In general terms, both the experimental feeding and strength distributions of the two decays show a prominent peak at low energy (around 1 MeV), which dominated the high resolution data. 

For the theoretical description the formalism described in \cite{Sarriguren2005,Moreno2006} was employed. It uses the quasiparticle-random-phase approximation (QRPA), with an axially deformed Hartree-Fock (HF) mean field generated by Skyrme forces, and pairing treated in the BCS framework. The equilibrium deformation is obtained self-consistently as the shape that minimises the energy. The potential energy curves obtained with this formalism exhibit oblate, spherical and prolate minima lying close in energy that are identified as the ground state and low-lying intruder states. 
Then the HF+BCS+QRPA approach is applied to the calculation of the B(GT) distribution, assuming that the parent and daughter nuclei have the same deformation. The $B$(GT) distributions of the $^{184-194}$Pb isotopes clearly show different patterns depending on the assumed deformation of the parent ground state ( Fig. 5 of \cite{Sarriguren2005} ), a feature that can be used for shape determination if combined with a proper determination of the beta strength distribution.
It is worth noting that while the relative equilibrium energies are sensitive to both 
Skyrme and pairing interactions \cite{Sarriguren2005}, the B(GT) distributions are more sensitive to the nuclear deformation than to those interactions \cite{Sarriguren2005}.
In the following discussion the theoretical results of \cite{Moreno2006} using a Sly4 Skyrme force are used. 
  
Although lead nuclei are expected to be spherical in their ground states \cite{Witte2007}, the use of this method can be viewed as a test to confirm its validity in this region, with the idea of  extending it to other possible cases of interest such as Pt, Po and Hg,  which are not necessarily spherical in their beta-decaying states. We remind the reader, that this method has only been applied so far in the $ A \sim 80$  region. With the availability of a reliable technique to measure the $B$(GT) strength over the whole $Q_{EC}$ window, the possibility is open also for further tests of nuclear models in the region around $Z = 82$.

\begin{figure}[b]
    \begin{center}
    \includegraphics[width=6.5cm]{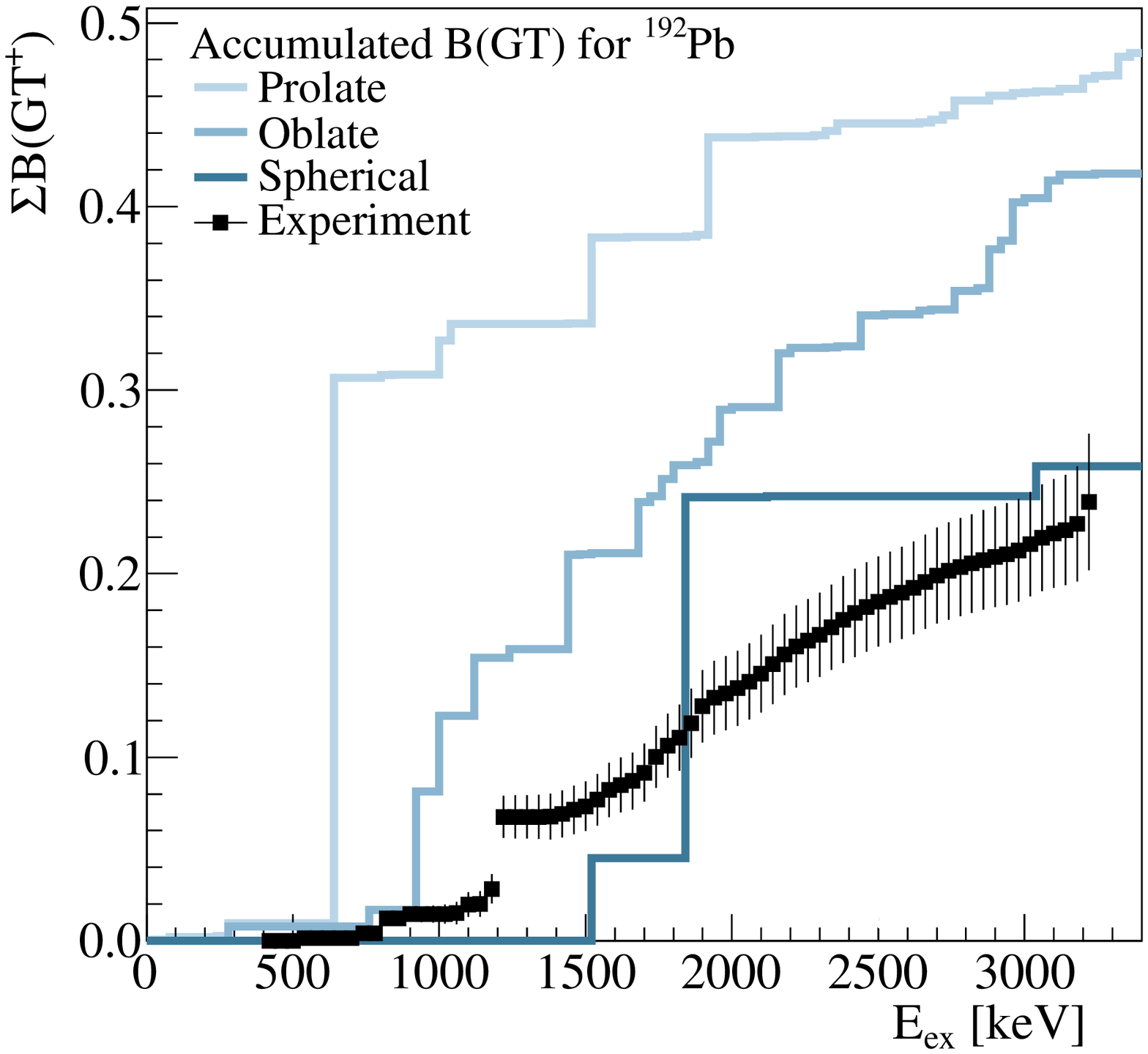}\hspace{5mm}\includegraphics[width=6.5cm]{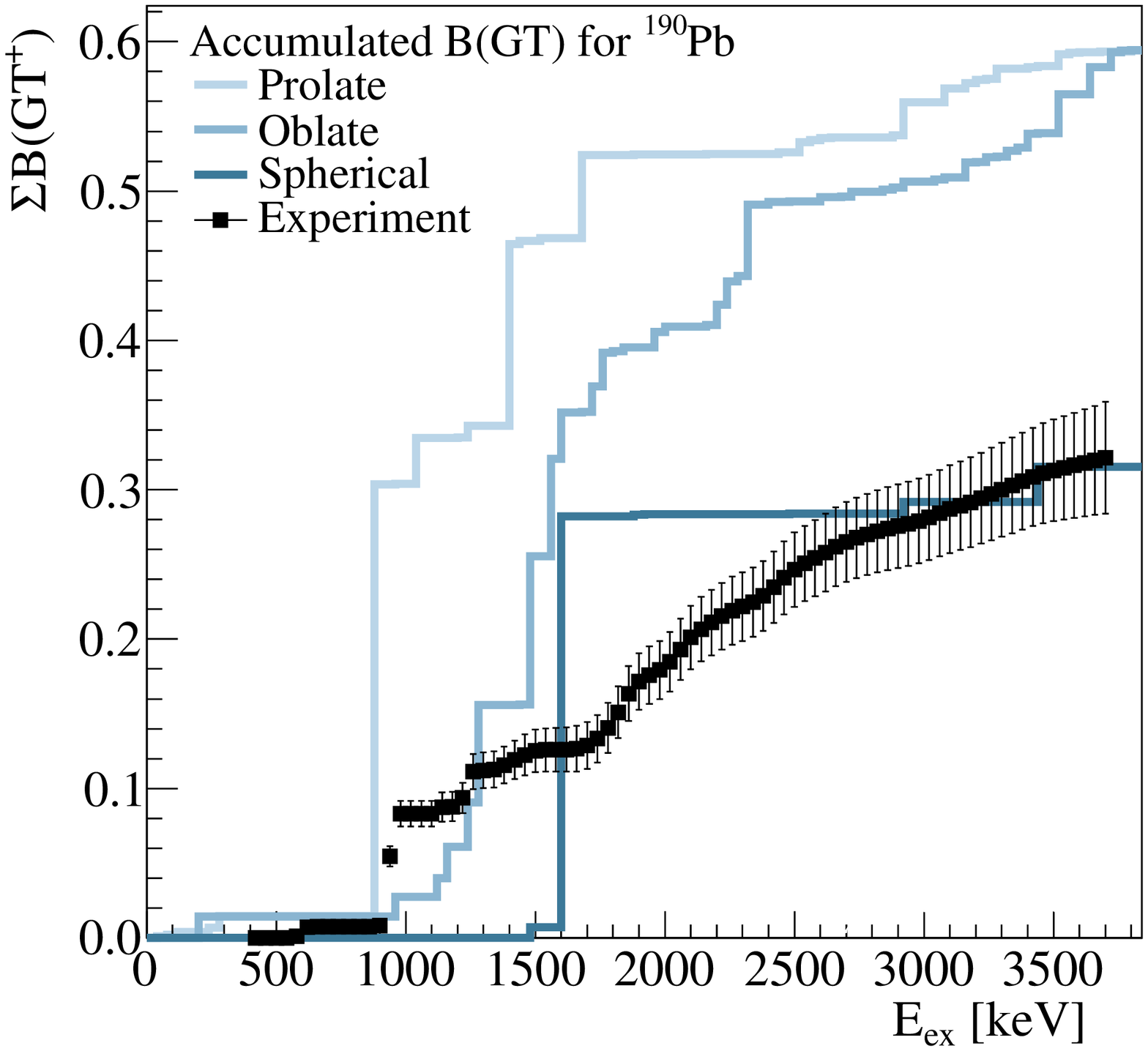}
    \caption{Comparison of the accumulated strength for the beta decays of $^{192,190}Pb$ with theory. The theoretical results are presented in two ways: the accumulated direct results from the calculations and accumulated strengths calculated with folded distributions using 1 MeV width Gaussians on the discrete spectrum. }\label{fig:acupers}
    \end{center}
  \end{figure}

One of the best ways to compare theory and experiment is using the accumulated beta strength. The accumulated strength  
at each energy was obtained by adding the strength observed up to that excitation energy. The results are presented in Fig. \ref{fig:acupers}. In the comparison the theoretical results from Ref. \cite{Moreno2006} have been scaled by a standard quenching factor $(g_A/g_V)_{eff} = 0.77(g_A/g_V)$, that is, a global factor of $0.77^2 \sim 0.6$. The theoretical calculations produce sizeable fragmentation of the strength only in the deformed cases, which is reflected by the ``staircase" character of the accumulated strength.  The two panels show a nice agreement with the theoretical results (total strength) for spherical ground states, but showing less spreading or fragmentation than in the $ A \sim 80$ region cases \cite{Nacher2004,Poirier2004,Perez2013}, a feature partially related to the spherical character of the nuclei studied. 
In particular the total B(GT) strength to levels below the $Q_{EC}$ window is very well reproduced. 

In the spherical case, if we compare the theoretical and experimental 
excitation energies where the strongest populated peak appears, we see an 
energy difference of approximately 0.7 MeV that is mostly related to 
the difficulties of these forces in reproducing  single-particle 
energies accurately. Skyrme interactions were designed to provide universal 
descriptions of nuclear properties over the whole nuclear chart and 
detailed spectroscopy is beyond the scope of the present approach.
When moving to more exotic neutron-deficient Pb isotopes the measurements 
indicate that the energy of the state where most of the strength is concentrated decreases
and the total strength increases (including the $^{188}$Pb decay case not presented here \cite{Algora188}). These general trends are well accounted for 
by the calculations. 



The quantitative description of mixed shapes is beyond the mean field approach used at present, and one should consider that the experimental results are in nice agreement with theory because of the expected small degree of mixing. Based on the present results we can infer  spherical character for the  $^{190,192}Pb$ ground states. This represents an additional and independent proof of the results obtained in \cite{Witte2007}, where small deviations from the liquid drop model for these nuclei were interpreted as the effect of tiny admixtures of intruder contributions in the ground state wave functions, but essentially they were considered to be spherical. Another conclusion of our work is that the method can be applied to heavy nuclei and suggests that  this technique (TAS measurements combined with theoretical calculations) is suitable for use with other deformed nuclei in the region such as the Hg and Pt isotopes, which are of particular interest because of the existence of isomeric states with different shapes. 

This work has been supported by the Spanish MINECO through projects
FPA2008-06419-C02-01, FPA2011-24553, FPA2013-41267-P, FPA2014-52823-C2-1-P, FRPA2012-32443, 
FIS2011-23565 and CSD-2007-00042 (CPAN Consolider, Ingenio2010), by the European 
Union Seventh Framework via ENSAR (contract no. 262010), contract PIOF-GA-2011-298364 and by OTKA contracts K100835 and K106035. 




\bibliography{article_prc_ale_3}







\end{document}